\documentclass[review]{elsarticle}

\usepackage{geometry}
\usepackage[colorlinks=green,
linkcolor=red,
anchorcolor=blue,
citecolor=green]
{hyperref}

\usepackage{times}
\usepackage{soul}
\usepackage{url}

\usepackage{booktabs}
\usepackage{times}
\usepackage{soul}
\usepackage{url}
\usepackage{amsmath}
\usepackage[utf8]{inputenc}
\usepackage{bbding}
\usepackage{graphicx}
\usepackage{amsmath}
\usepackage{graphbox}
\usepackage{booktabs}
\usepackage{algorithm}
\usepackage{algorithmic}
\usepackage{pifont}
\usepackage{framed}
\usepackage{graphicx}
\usepackage{subfigure}
\usepackage{multirow}
\usepackage{amsmath}
\usepackage{amsfonts}
\usepackage{color}

\emergencystretch=\maxdimen
\journal{Journal of Computers \& Security}

\begin{document}
\begin{frontmatter}
\title{AdvCheck: Characterizing Adversarial Examples\\ via Local Gradient Checking}

\author[mymainaddress]{Ruoxi~Chen}
\ead{2112003149@zjut.edu.cn}
\author[mymainaddress]{Haibo Jin}
\ead{2112003035@zjut.edu.cn}

\author[mymainaddress,mysecondaryaddress]{Jinyin~Chen\corref{mycorrespondingauthor}}
\cortext[mycorrespondingauthor]{Corresponding author}
\ead{chenjinyin@zjut.edu.cn}
\author[mymainaddress,mysecondaryaddress]{Haibin~Zheng}
\ead{haibinzheng320@gmail.com}

\address[mymainaddress]{College of Information Engineering, Zhejiang University of Technology, Hangzhou, 310023, China}
\address[mysecondaryaddress]{Institute of Cyberspace Security, Zhejiang University of Technology, Hangzhou, 310023, China}


\begin{abstract}
Deep neural networks (DNNs) are vulnerable to adversarial examples, which may lead to catastrophe in security-critical domains. Numerous detection methods are proposed to characterize the feature uniqueness of adversarial examples, or to distinguish DNN's behavior activated by the adversarial examples. Detections based on features cannot handle adversarial examples with large perturbations. Besides, they require a large amount of specific adversarial examples. Another mainstream, model-based detections, which characterize input properties by model behaviors, suffer from heavy computation cost. To address the issues, we introduce the concept of local gradient, and reveal that adversarial examples have a quite larger bound of local gradient than the benign ones. Inspired by the observation, we leverage local gradient for detecting adversarial examples, and propose a general framework AdvCheck. Specifically, by calculating the local gradient from a few benign examples and noise-added misclassified examples to train a detector, adversarial examples and even misclassified natural inputs can be precisely distinguished from benign ones. Through extensive experiments, we have validated the AdvCheck's superior performance to the state-of-the-art (SOTA) baselines, with detection rate ($\sim \times 1.2$) on general adversarial attacks and ($\sim \times 1.4$) on misclassified natural inputs on average, with average 1/500 time cost. We also provide interpretable results for successful detection. 
\end{abstract}

\end{frontmatter}

\section{Introduction}
Thanks to the outstanding performance and stable capability, deep neural networks (DNNs) are now widely used in various areas, e.g., image classification~\cite{zhang2021source}, object detection~\cite{zhang2022misleading}, malware detection~\cite{qiang2022efficient} and behavior classification~\cite{chen2022efficient}. However, DNNs may expose incorrect predictions due to subtle changes of the input, especially on adversarial perturbations. Their undesirable vulnerabilities of security and robustness problems have become a major security concern. These issues will pose grave threat, especially when DNNs are deployed in the safety-critical domains like autopilot~\cite{eykholt2018robust} and identity recognition~\cite{rozsa2019facial}.

Numerous adversarial attack methods have been proposed from various perspectives, fooling DNNs with different perturbations. They can be classified into white-box attacks and black-box attacks, according to the knowledge of the targeted model. Typical white-box attacks mainly use gradient information to determine misleading perturbations~\cite{GoodfellowSS14,Kurakin2017Adversarial,papernot2016limitations,Madry2018Towards,dong2018boosting}. The structure and parameters of the targeted model should be given in advance before implementation of white-box attacks. On the contrary, black-box attacks mislead the targeted model only with the output of it. They are mainly decision-based and score-based~\cite{Brendel2018Boundary,SchottRBB19,su2019one}. Compared with white-box attacks, perturbations of black-box attacks usually have larger size.

To counter these attacks, far intensive efforts have been made to identify security-compromised input sources, following two design goals: (1) high detection rate; (2) low computation cost and time complexity. 
Existing detection methods are brought up 
mainly based on the feature difference of adversarial examples and benign examples, or based on the model behavior difference activated by them. 
Image pre-processing based methods~\cite{Xu0Q18,LiangLSLSW21}, as one of the feature-based methods, introduced image transformations to the suspicious input, and attempted to detect adversarial examples by comparing the classification results before and after operations. But they don't perform well on larger perturbations and their effects rely heavily on properly-set parameters.
Another family of feature-based detections are designed from distance and neighbors in in the feature representation~\cite{Ma0WEWSSHB18,CohenSG20,yang2020ml}. They computed distance metrics of each layer in the DNN and characterized key properties of the adversarial subspace. However, to achieve good performance against general attacks, they require a large amount of different adversarial examples and long time for feature extraction, which increases the data dependency and computation cost.

When the model is fed with adversarial examples and benign ones, it's more straightforward and general to address the problem of adversary detection from the aspect of the model's different behaviors, i.e., neuron activations. As model-based method, Ma et al.~\cite{ma2019nic} extracted value invariants and the provenance invariants from benign training data, and then used them for adversarial detection. The proposed method NIC, traverses each neuron in each layer, so it suffers from high computation burden, e.g., for VGG19~\cite{simonyan2014very} with 50,782 neurons, the times of calculation is over $5e^4$. 

Recall that the decision results of DNNs are determined by the aggregation of each channel, benign and adversarial examples activate different values of channels \cite{ma2019nic, bai2020improving}. So we try to find the difference in a certain layer in the DNN through feeding benign and their corresponding adversarial examples. Since activation values in a certain layer are often normalized to [0,1], they don't show significant difference before and after attacks. To magnify the difference, we try to take the partial derivative of the output of the layer and propose the concept of \emph{local gradient}, which quantitatively measures the extent of input's contribution to layer's activation values.

\begin{figure}[t]
\centering
        \includegraphics[width=0.65\linewidth]{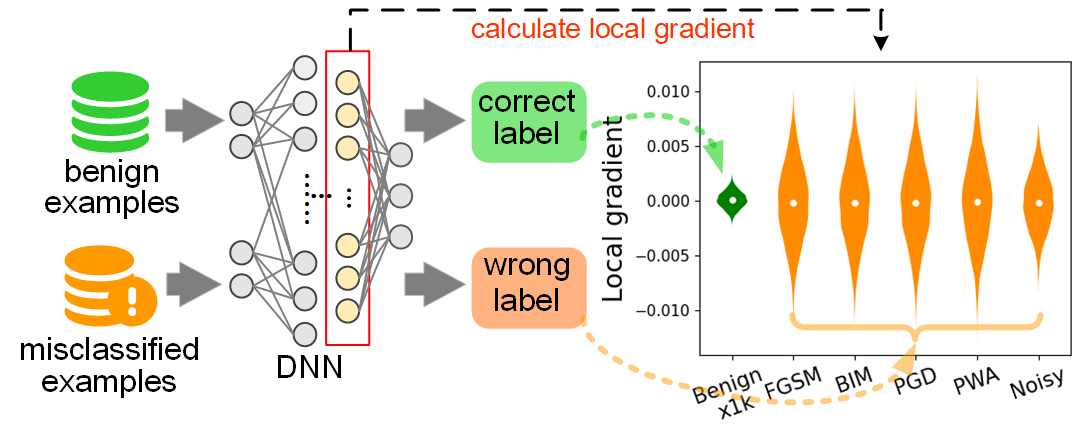}
\caption{The distribution of local gradient of benign and misclassified examples. The violin plot shows the data distribution shape and its probability density and the white point is the median.}
	\label{intro}
\end{figure}

To further observe the difference of local gradient before and after attacks, we compute local gradient between benign and adversarial examples, and then visualize the distribution of it in Fig. \ref{intro}, where green and orange violins denote the distribution of local gradient of benign and misclassified examples, respectively. Here, the dropout\_2 layer in VGG19 of CIFAR-10~\cite{krizhevsky2009learning} is used for visualization. Values of benign examples are magnified 1000 times for better visualization. 500 adversarial examples from fast gradient sign method (FGSM)~\cite{GoodfellowSS14}, basic iterative method (BIM)~\cite{Kurakin2017Adversarial}, projected gradient descent (PGD)~\cite{Madry2018Towards}, point wise attack (PWA)~\cite{SchottRBB19} and misclassified examples by adding random noise (dubbed as ``Noisy'') are fed into the model. It can be easily observed that the local gradient of adversarial examples have quite larger bound than that of benign ones. Specifically, adversarial examples generated by different algorithms, as well as noisy examples, show significantly large bound value of the local gradient, i.e., absolute values up to a thousand times of benign ones. This illustrates that large difference in local gradient between benign and adversarial examples can be used to characterize the properties of adversarial examples.

Inspired by this observation, we leverage the local gradient for adversarial detection and further design a general method, namely AdvCheck, against general adversarial examples. First, we calculate the local gradient in a certain layer in the DNN by a few batches of benign and perturbed examples. To generate these perturbed examples, random noises are added on benign examples until they are misclassified. Thus, their local gradient is used to simulate that of the adversarial examples. Then an adversarial detector will be trained to learn the distributions of the local gradient. The well-trained detector can precisely characterize the properties of the input example. Based on general pattern of adversarial examples in the local gradient, AdvCheck does not require explicit prior knowledge of specific attacks. It achieves attack-agnostic detection and can be served as an effective tool against general adversarial examples. 

Our main contributions are summarized as follows:
\begin{itemize}
    \item We propose the concept of local gradient, to measure the extent of input's contribution to layer's activation values. We observe that adversarial examples show a quite larger bound of local gradient than that of benign ones. 
    \item Based on the observation, we design a detection method AdvCheck for differentiating benign and adversarial examples. Only relying on a few benign examples, AdvCheck can precisely detect general adversarial examples, and even misclassified natural inputs.
    \item Extensive experiments on benchmark datasets have been conducted to verify AdvCheck's performance. Results have shown that AdvCheck is superior to the state-of-the-art (SOTA) detection methods against general adversarial examples.
\end{itemize}

\section{Related Work}
In this section, we review the related work and briefly introduce adversarial attacks and detection methods against them. 

\subsection{Adversarial Attacks}
According to whether the attacker requires prior knowledge of the targeted model, adversarial attacks can be divided into white-box attacks and black-box attacks.

\subsubsection{White-box Attacks}
Based on full knowledge of model structure and parameters, white-box attacks are easy to conduct with small perturbations. Goodfellow et al.~\cite{GoodfellowSS14} proposed FGSM, adding perturbations on the direction where the gradient of the model changes most. Based on it,  Kurakin et al.~\cite{Kurakin2017Adversarial} expanded the operation on loss function into several small steps and proposed BIM. Dong et al. ~\cite{dong2018boosting} introduced momentum term into the iterative process and they proposed momentum-based iterative FGSM (MI-FGSM). Adversarial examples generated by it are more transferable towards models. Similarly, PGD~\cite{Madry2018Towards} achieves stronger attack by projecting the perturbation to the specified range iteratively. Papernot et al. \cite{papernot2016limitations} leveraged critical pixels in the saliency map and designed Jacobian-based saliency map attack (JSMA) method. Moosavi et al.~\cite{moosavi2016deepfool} gradually pushed the image within the classification boundary to the outside until the wrong predicted label occurred. Besides, Modas et al.~\cite{modas2019sparsefool} proposed SparseFool based on the low mean curvature of the decision boundary to controls the sparsity of the perturbations. Duan et al. \cite{duan2021mask} combined the rotation input strategy with iterative attack method to generate stronger adversarial images with more imperceptible adversarial noises. Based on attention mechanism, Chen et al.~\cite{chen2021finefool} focused more on object contours and proposed FineFool, which achieves a higher ASR with smaller perturbations. 

Besides, Croce et al. \cite{croce2020reliable} combined extensions of PGD attack with other complementary attacks. They proposed AutoAttack, which is parameter-free, computationally affordable and user-independent to test adversarial robustness.

\subsubsection{Black-box Attacks}
Different from white-box attacks, black-box ones fool the model by the model confidence or output labels. Brendel et al.~\cite{Brendel2018Boundary} sought smaller perturbations based on the model decision while staying adversarial. PWA attack~\cite{SchottRBB19} starts with an adversarial and performs a binary search between the adversarial and the original until the input is misclassified. Similarly, additive uniform noise attack (AUNA)~\cite{FoolboxAUNA} adds uniform noise to the input until the model is fooled. Andriushchenko et al.~\cite{andriushchenko2020square} designed Square attack based on random search, which achieves higher success rate in untargeted setting with query-efficiency. These black-box attacks are limited when fooling DNNs tailored for high-resolution images. To solve this problem, Wang et al.~\cite{wang2022pisa} proposed a pixel skipping and evolutionary algorithm-based attentional black-box adversarial attack PISA in attacking high-resolution images.

\subsection{Adversarial Detections}
Adversarial detectors can distinguish adversarial examples from benign ones, which serves as an alarming bell in the system. They are developed along two mainstreams: feature-based and model-based detections.

\subsubsection{Feature-based Detections}
This type methods distinguish adversarial examples by characterizing features of input examples. Early-proposed detections are mainly based on image pre-processing. They achieve detections by comparing the difference before and after image transformations. Meng et al.~\cite{Meng2017MagNet} trained a denoising encoder to reconstruct the image, and proposed MagNet for detection. Xu et al.~\cite{Xu0Q18} conducted reduction of color bit depth and spatial smoothing to adversarial examples. Detection results are based on the difference between the original input and the pre-processing input.  Tian et al.~\cite{tian2018detecting} found that adversarial images are sensitive to transformation operations, while benign ones are not. They implemented 45 different image transformation methods, achieving a fair good detection ratio. Liang et al.~\cite{LiangLSLSW21} introduced scalar quantization and smoothing spatial filter for implementing an adaptive noise reduction (ANR) to adversarial examples. These methods are easy to implement but may be compromised when faced with larger adversarial perturbations.

Another family of these methods are based on feature representation of models. Ma et al. \cite{Ma0WEWSSHB18} characterized dimensional properties of adversarial regions by local intrinsic dimentionality (LID), based on the distance distribution of the example to its neighbors in the feature subspace. Papernot et al. \cite{papernot2018deep} introduced the deep k-nearest neighbors (DkNN) to estimate the given input. By comparing it to its neighboring training points based on distance of feature representations learned in each layer, DkNN can be applied to detect adversarial examples. Following them, Cohen et al. \cite{CohenSG20} trained an adversarial detector and proposed NNIF for distinguishing adversarial attacks, which is based on influence function and k-nearest neighbor ranks in the feature space. Yang et al. \cite{yang2020ml} observed that the feature attribution of adversarial
examples near the boundary always differs from those of corresponding benign examples. They further designed ML-LOO through thresholding a scale estimate of feature attribution scores. Such approaches based on feature representations show limitations in computation efficiency, i.e., they require long time for extracting features from multiple layers to train an extra adversarial detector.

\subsubsection{Model-based Detections}
Different from feature-based methods, model-based detections focus on model behaviors. Wang et al. \cite{WangD00Z19} found that adversarial samples are much more sensitive than benign samples on mutation of DNNs. They then integrated statistical hypothesis testing and model mutation testing to check the property of the input. However, based on model mutation, this methods requires heavy computation resources. 

On the other hand, Ma et al. \cite{ma2019nic} attempted to identify adversarial examples through comparing commonly-activated neurons of adversarial examples. The proposed method, NIC, uses neuron activations from all layer to calculate value invariants and provenance invariants for detection. It requires a huge overhead of parameters or runtime since it has to traverse all neurons in the model. Different from NIC, we focus on abnormal behaviors of the model in one layer. In this sense, our method reduces the complexity of detection. Inspired by neuron coverage in software testing, Sperl et al.~\cite{SperlKCLB20} extracted the activation values of all available dense layers as dense layer neuron coverage. They leveraged the difference in it between benign and adversarial examples, and proposed dense-layer-analysis (DLA) for detection. But DLA relies on large amount of adversarial examples and large memory for holding activation value sequences of one specific attack.

Lust et al. \cite{lust2020gran} focused on model's gradient and assumed that misclassified examples have a larger gradient than correctly classified examples. They proposed a parameter-efficient detection method GraN, which calculates model loss of the input and its smoothing one. The model loss is then transformed into feature vector of each layer and the final detection output is given by logistic regression network. Different from them, we aim to spot adversarial characters in one layer for detection, not the model loss. The detector of AdvCheck is based on fully-connected layers while GraN uses logistic regression network for predicting model misclassification probability.

\section{Preliminaries}
In this section, we define threat model in AdvCheck and give definitions of local gradient. We also give mathematical analysis of local gradient and adversarial attacks.

\subsection{Threat Model}
A DNN-based classifier can be denoted as $f(x): X\rightarrow Y$, where $x \in X \subset \mathbb{R}^N$ represents the input and $Y$ denotes predicted classes. We use $y$ to represent ground truth label of $x$. For a well-trained model $f(x)$, the prediction can be formulated as $f(x)=y$. 

Attackers attempt to craft small perturbations added on the benign example, the adversarial example $x_{adv}$ of $x$ is calculated as $x_{adv}=x+\Delta x$. The perturbation $\Delta x$ is constrained by $\epsilon$, i.e., $ ||\Delta x||_{p} \leq \epsilon$, where $\epsilon > 0$, $p \in \mathbb{N}$, and $||\cdot||_{p}$ represents the $L_{p}$-norm distance. If $f(x_{adv})\neq f(x)$, the adversarial example $x_{adv}$ is considered to successfully attack the model.

We define two scenarios of attackers' knowledge. For white-box attacks, attackers have full knowledge of the targeted model, e.g., the model structure, weights and prediction probability.  For the black-box scenario, attackers only know model confidence or predictions, or rely on queries to optimize adversarial perturbations. 

Our goal is to detect adversarial attacks, i.e, find adversarial examples in advance. We don't know the specific type of attacks, regardless of white-box or black-box ones. But we can access part of correctly-classified benign examples with their corresponding labels from the training dataset. 

\subsection{Definitions\label{local_def}}
The DNN consists of multiple layers: the input layer, many hidden layers and the output layer. Each layer consists of multiple neurons. The output of the layer is the nonlinear combination of each neuron’s activation state in the previous layer.

Given a DNN and let $X=\{x_1, x_2, ...\}$ denotes a set of inputs. For $m$ class labels, $f(x)=\{y_1(x), y_2(x), ..., y_m(x)\} \in \mathbb{R}^m$ denotes the confidence values before softmax. The predicted label $c$ is given as: $c=\text{argmax}_{1\leq i\leq m}f(x)$. We denote the corresponding confidence of the predicted label $c$ as $y_c$.

\textbf{Definition 1: Layer Output.} Let $N_i=\{n_{i,1},n_{i,2}, ...\}$ denotes the set of neurons in the $i$-th layer. We define the layer output $\varphi_i(x)$ of the $i$-th layer as:
\begin{equation}
\varphi_i(x)=\bigcup_{i=j}^{a}\psi_{i,j}(x)
    \label{Layer-wise Output}
\end{equation}
where function $\psi_{i,j}(\cdot)$ represents the output of the $j$-th neuron in the set $N_{i}$, when fed with input $x \in X$. $a$ is the total number of neurons in that layer.

The output of one layer is served as the input for the next layer. So the output of the $(i+1)$-th layer can be calculated as:
\begin{equation}
    \varphi_{i+1}(x)=\sigma(\boldsymbol{\omega_{i}}~\varphi_i(x)+\boldsymbol{b_{i}})
\end{equation}
where $\boldsymbol{\omega_{i}(\cdot)}$ and $\boldsymbol{b_{i}}$ denote the weight and bias values of $i$-th layer, respectively. $\sigma$ is the activation function and for brevity, we omit it in the following paper.

\textbf{Definition 2: Local Gradient.} Local gradient measures the extent of input's contribution to the output in a certain layer. For the given DNN, the local gradient $\mu$ of the input $x$ on the neuron $n$ in the $i$-th layer is defined as: 
\begin{equation}
    \mu_i(x)=\frac{\partial y^c}{\partial \varphi_i(x)}
    \label{local_gradient}
\end{equation}
where $y^c$ denotes the confidence of predicted label $c$ when fed with input $x$. $\partial$ denotes partial derivative function. It must be noted that different from model gradient used to craft adversarial perturbations, local gradient takes partial derivatives for a certain layer, instead of the input.


\subsection{Local Gradient And Adversarial Attacks\label{emp_study}}
We interpret the essence of adversarial attacks from the novel perspective of layers, i.e., adversarial examples lead to output changes of models due to accumulations of deviations on layer's output.

The decision result of DNN is determined by the nonlinear combination of each layer along forward propagation. The DNN is thus a combination of composite function, formulated as follows:
\begin{equation}
    f(x)=\varphi_l \circ \varphi_{l-1} \circ \cdots \circ \varphi_2 \circ \varphi_1(x)
\end{equation}
where $l$ is the total number of layers in the model. Specifically, $\varphi_1$ is the input layer and $\varphi_l$ is the output layer. 

Recall that the adversarial example $x_{adv}$ is calculated by adding small perturbations to the input, i.e., $x_{adv}=x+\Delta x$. When the DNN is fed with $x_{adv}$, the output of each layer will be changed. Take the first layer as an example. In that layer, $\varphi_1(x)=\boldsymbol{\omega_{1}}(x) \cdot x+\boldsymbol{b_1}$. The weight in that layer can be regarded as: $\boldsymbol{\omega_1}(x+\Delta x)=\boldsymbol{\omega_1}(x)+ \Delta \boldsymbol{\omega}$. We calculate the output of the first layer as follows:
\begin{equation}
\begin{aligned}
    \varphi_1(x_{adv})=&\boldsymbol{\omega_1}(x+\Delta x) \cdot (x+\Delta x)+\boldsymbol b_1\\
    =& \boldsymbol{\omega_1}(x)\cdot x+\boldsymbol{\omega_1}(x) \cdot \Delta x+ \Delta \boldsymbol{\omega} \cdot x+\Delta \boldsymbol{\omega} \cdot \Delta x +\boldsymbol b_1\\
    =& \varphi_1(x) + \underbrace{\boldsymbol{\omega_1}(x) \cdot \Delta x+ \Delta \boldsymbol{\omega} \cdot x+\Delta \boldsymbol{\omega} \cdot \Delta x}_{\Delta \varphi_1(x, \Delta x)}+\boldsymbol b_1 
\end{aligned}
\end{equation}
where $\varphi_{1}(x_{adv})$ and $\varphi_{1}(x)$ denote the output of the first layer when the model is fed with the adversarial and benign example, respectively. From the equation, the output of this layer will be affected by last three items in the formula and we sum them as $\Delta \varphi_1(x, \Delta x)$. Such small changes will be magnified during the forward propagation between layers, we formulate this process as:
\begin{equation}
    \begin{aligned}
    f(x_{adv})=&\varphi_l \circ \varphi_{l-1} \circ \cdots \circ \varphi_2 \circ \varphi_1(x_{adv})\\
    =&\varphi_l \circ \varphi_{l-1} \circ \cdots \circ \varphi_2 \circ (\varphi_1(x)+\Delta \varphi_1(x, \Delta x))    
    \end{aligned}
\end{equation}
The predicted label is given as $c_{adv}=\text{argmax}_{1\leq i\leq m}f(x_{adv}) \neq c$. Adversarial perturbations finally lead to changes of predicted confidence, i.e., misclassification.

Equipped with local gradient, we ask the following question: Does local gradient behave differently on adversarial examples and benign inputs?
If the difference exists between benign and adversarial inputs, we can then possibly separate benign inputs from adversarial ones. 
\begin{figure}[htbp]
\centering
    \subfigure[conv2d\_2]{
        \includegraphics[width=0.32\linewidth]{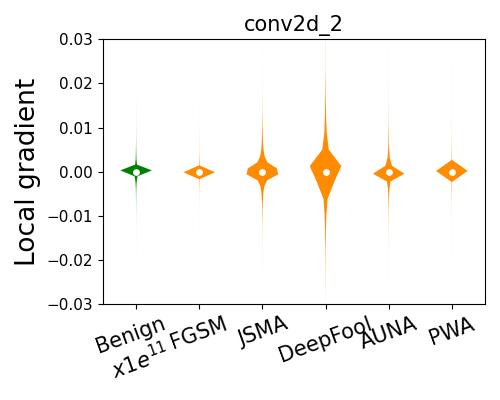}}
    \subfigure[max\_pooling2d\_2]{
        \includegraphics[width=0.32\linewidth]{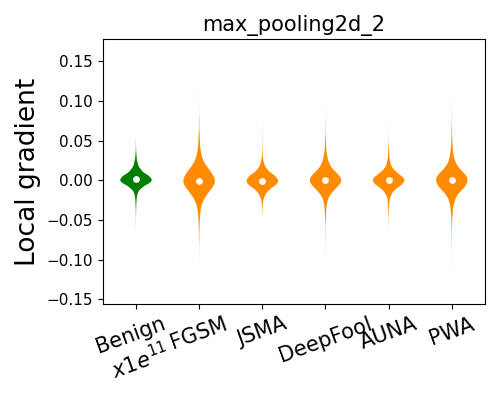}}
    \subfigure[dense\_2]{
        \includegraphics[width=0.32\linewidth]{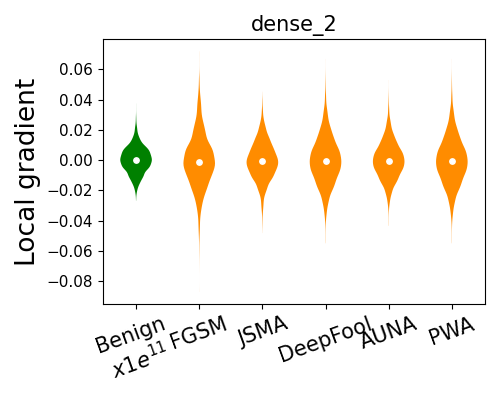}}\\
    \subfigure[conv2d\_13]{
        \includegraphics[width=0.32\linewidth]{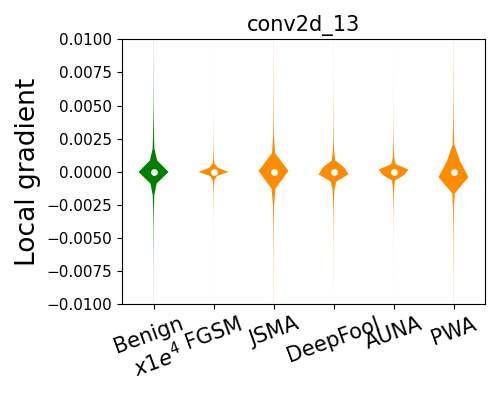}}
    \subfigure[BN16]{
        \includegraphics[width=0.32\linewidth]{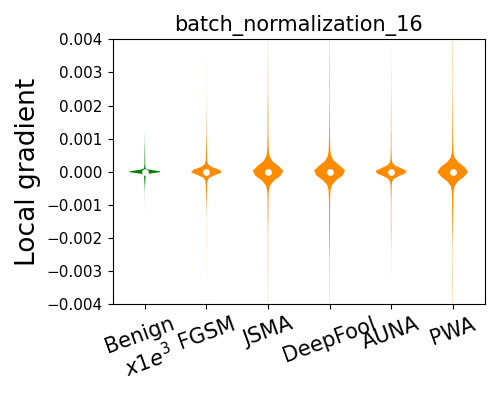}}
    \subfigure[add\_8]{
        \includegraphics[width=0.32\linewidth]{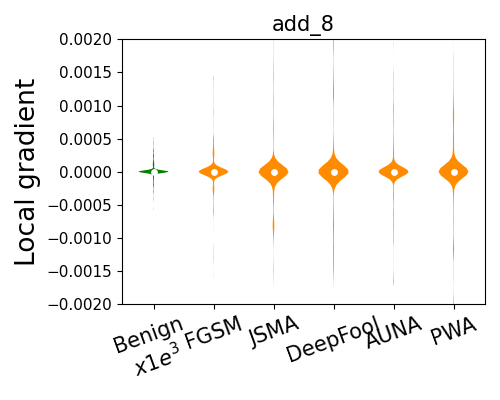}}
\caption{Local gradient of benign and adversarial examples. The first row are from AlexNet  of CIFAR-10 while the second row from ResNet20 of GTSRB.}
\label{emp}
\end{figure}

We experimentally study the difference between benign and adversarial examples. Specifically, we show that, when we choose different layers in different models, the bound of local gradient of adversarial examples are significantly larger than that of benign ones. We conduct experiments with
various adversarial attacks including FGSM~\cite{GoodfellowSS14}, JSMA~\cite{papernot2016limitations}, DeepFool~\cite{moosavi2016deepfool}, AUNA~\cite{FoolboxAUNA} and PWA~\cite{SchottRBB19}. The results are shown in Fig. \ref{emp}, where the first row are from AlexNet~\cite{krizhevsky2012imagenet} of CIFAR-10~\cite{krizhevsky2009learning}, the second row from ResNet20~\cite{he2016deep} of GTSRB~\cite{Stallkamp-IJCNN-2011}. ``BN" denotes batch normalization. The violin plot shows the data distribution shape and its probability density, which combines the features of box and density charts. The white point is the median. Values of benign examples are magnified for better visualizations and the magnification is reported in the x-ticks.

As we can see, the value bound of local gradient of adversarial examples in different layers are significantly larger than those of benign examples, for all the adversarial attacks, regardless of white-box or black-box attacks.

\section{AdvCheck}
Based on the above observation, we leverage the local gradient for detecting adversarial examples. In this part, we describe AdvCheck in detail. We further analyze the algorithm complexity of AdvCheck and compare it with other methods.

\subsection{Overview}

\begin{figure}[htbp]
\centering
        \includegraphics[width=0.62\linewidth]{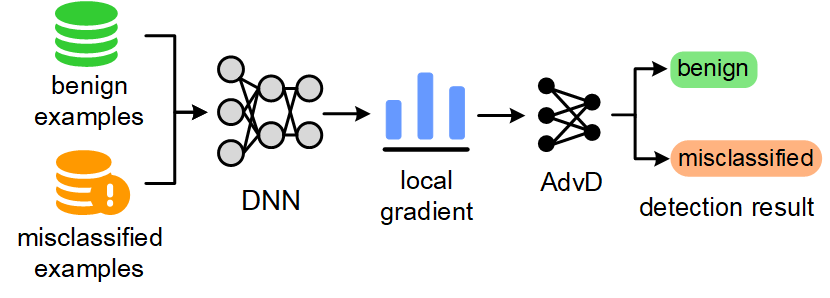}
\caption{The workflow of AdvCheck, including calculating the local gradient and training the detector AdvD.}
	\label{overview}
\end{figure}

The workflow of AdvCheck is shown in Fig. \ref{overview}. AdvCheck consists of two major components: calculating local gradient and training an adversarial detector AdvD. To begin with, benign and misclassified examples are input to the DNN, respectively. The corresponding local gradient in the chosen layer will be calculated and further used to train the AdvD. Finally, the well-trained AdvD can correctly distinguish benign and misclassified examples. 

\subsection{Calculation of Local Gradient}
In this step, we calculate local gradient that will be further used for training the AdvD. 

We added random noise towards benign examples until they are misclassified. Then benign examples and their corresponding misclassified examples are fed into the model for calculating local gradient. The maximum perturbation size of the input examples is limited by 0.25 in $l$-2 norm. We use random noise-added examples here, instead of adversarial examples of specific attacks. We assume that random noise will be more general to cover various adversarial perturbations. Besides, it is easy to craft with high efficiency.

In this process, in the default setting, the layer between convolutional and dense layer is chosen, e.g., flatten layer or global average pooling.
These layers contain both pixel features and high-dimensional features, playing decisive role in classification. The impact of chosen layer will be discussed in the experiment.

\subsection{Detection via AdvD}
As illustrated in Section~\ref{local_def}, local gradient of benign and misclassified examples show large difference. So based on it, we design the adversarial example detector (AdvD) to determine whether the input data is adversarial.

For each targeted model, we train one detector. AdvD consists of several fully-connected layers, whose detailed structures will be stated in the experiment. The input layer size of it is the same shape as the local gradient calculated in the previous step. The local gradient of benign and misclassified examples will be concatenated to generate the training set of AdvD. The input of AdvD is the concatenated local gradient, and the output is the detection result. ``benign" means benign example, marked with 0, and ``misclassified" means misclassified example, marked with 1. During the training process, the loss function of AdvD is defined as follows: 
\begin{equation}
\begin{aligned}
    loss_{\rm AdvD}= CE(\text{AdvD}&(F(s)),y), y \in \{0,1\}\\
    F(s) \in \{Concat(\mu_{i}&(x),~\mu_{i}(x^{*}))\}\\
    \end{aligned}
\end{equation}
where $\rm AdvD(\cdot)$ represents the output of AdvD. $Concat(\cdot)$ is concatenate function. CE is the cross entropy.

The parameters of AdvD are updated by minimizing the loss function. By properly setting training parameters, AdvD can easily detect misclassified examples after training. 

The pseudo-code of AdvCheck is presented in Algorithm 1.

\makeatletter
\makeatother
\begin{figure}[htbp]

  \renewcommand{\algorithmicrequire}{\textbf{Input:}}
  \renewcommand{\algorithmicensure}{\textbf{Output:}}
  \begin{algorithm}[H]
    \caption{AdvCheck}
    \begin{algorithmic}[1]
      \REQUIRE Benign examples $X=\{x_1,x_2,…\}$. Noise-added misclassified examples ${X}^{*}=\{x_1^*,x_2^*,…\}$. The $i$-th layer with its output $\varphi$ in the given DNN. 
      \ENSURE AdvD 
      \STATE ${{\mu}_i(X)=\{\emptyset\},{\mu}_i(X^{*})=\{\emptyset\}}$
      \FOR {$x$ in $X$}
      \STATE {calculate $\mu_i(x)$ according to Equation \ref{local_gradient}} \\
      ${\mu}_i(X)$.append($\mu_i(x)$)
      \ENDFOR
      \FOR {$x^*$ in $X^*$}
      \STATE {calculate $\mu_i(x^*)$ according to Equation \ref{local_gradient}} \\
      $\mu_i(X^*)$.append($\mu_i(x^*)$)
      \ENDFOR
      \STATE{AdvD = train a classifier on (${\mu}_i(X)$, $\mu_i(X^*)$)}
    \end{algorithmic}
  \end{algorithm}
\end{figure}

\subsection{Time Complexity Analysis and Comparison}
Here we analyze the algorithm complexity of AdvCheck. 

Since AdvCheck needs to extract the local gradient for detection, so the time complexity can be calculated as: $T(\text{AdvCheck})\sim \mathcal{O}(m)$, where $m$ denotes the number of examples to be detected.


As for existing detections, LID needs to compute feature distance in each layer, so the running time is magnified by the total number of layers $l$. So its time complexity is $T(\text{LID})\sim \mathcal{O}(m \times l)$. On the contrary, ANR only conducts simple transformations towards adversarial examples, the computation complexity is $T(\text{ANR}) \sim \mathcal{O}(m)$. NNIF transverses all examples in the training dataset, so it requires long time for computation: $T(\text{NNIF})\sim \mathcal{O}(m \times t \times l) $, where $t$ denotes the number of examples in the training dataset. NIC has to traverse all neurons inside the model, time complexity of it is $T(\text{NIC})\sim \mathcal{O}(m \times n) $, where $n$ denotes the total number of neurons in the DNN. 

\begin{table}[htbp]
\centering
\caption{Comparison with other detections.}
\resizebox{0.5\linewidth}{!}{
\begin{tabular}{cccc}
\toprule
\textbf{Detection} & \textbf{Adv-free} & \textbf{Time complexity} & \textbf{Estimated} \\ \hline
ANR \cite{LiangLSLSW21}     & \Checkmark     & $\mathcal{O}(m)$                 & 0.05s             \\
LID \cite{Ma0WEWSSHB18}     & \XSolid     & $\mathcal{O}(m \times l) $             & 1.35s             \\
NIC \cite{ma2019nic}    & \XSolid         & $\mathcal{O}(m\times n)$               & 102.09s            \\
NNIF \cite{CohenSG20}     & \XSolid       & $\mathcal{O}(m \times t \times l) $            & 695.08s            \\
ML-LOO \cite{yang2020ml}  & \XSolid  & $\mathcal{O}(m \times l) $  & 225.40s\\
AdvCheck           & \Checkmark       & $\mathcal{O}(m) $                & 0.35s            \\ 
\bottomrule
  \label{comparison}
  \end{tabular}}
  \end{table}

We summarize the comparison of time complexity in in Table \ref{comparison}, where ``Adv-free" means adversarial examples are not required for the method. Regarding the computational cost, $m$ and $t$ denotes the number of examples to be detected and the number of examples in the training dataset, respectively. $n$ denotes the total number of neurons in the DNN while $l$ is the number of layers. We also evaluate the average time of detection per image on VGG19 of CIFAR-10, as shown in the ``Estimated" column. Among all detections, AdvCheck requires a relatively low complexity. Consistent with our analysis, AdvCheck has acceptable time cost, with only 1/4 and 1/290 that of LID and NIC. This is because AdvCheck relies on only one layer of the model knowledge, more simple than feature space based detections. But compared with simple image transformation like ANR, AdvCheck shows disadvantage, because it involves the  structure of the targeted model.

\section{Experiments And Analysis}
To demonstrate the performance of AdvCheck, extensive experiments have been carried out, including the following parts: 
\begin{itemize}
    \item \textbf{RQ1}: Does AdvCheck show superior performance in detecting various adversarial attacks?  
    \item \textbf{RQ2}: Can AdvCheck effectively detect misclassified natural inputs?
    \item \textbf{RQ3}: How is the reliability of AdvCheck when faced with adaptive attacks? 
    \item \textbf{RQ4}: How different hyper-parameters (e.g., training examples and chosen layer) impact AdvCheck?
\end{itemize}

\subsection{Experiment Setup\label{setup}}

\textbf{Datasets:} We evaluate the detection effectiveness of AdvCheck on CIFAR-10~\cite{krizhevsky2009learning}, GTSRB~\cite{Stallkamp-IJCNN-2011} and ImageNet~\cite{russakovsky2015imagenet}. CIFAR-10 contains 60,000 32$\times$32 RGB images, belonging to 10 and 100 classes. In our experiment, 50,000 examples are used for training and 10,000 for validation set. GTSRB consists of more than 50,000 48$\times$48 German traffic sign images from 43 classes. We split the original training dataset into two parts, i.e., 70\% for training set and 30\% for validation set. For ImageNet, 11,000 images are selected for training and 2,000 images for testing.

\textbf{DNNs:} We adopt various classifiers on several benchmark datasets. For CIFAR-10 dataset, VGG19~\cite{simonyan2014very} and AlexNet~\cite{krizhevsky2012imagenet} models are adopted. For GTSRB, We train LeNet-5~\cite{lecun2015lenet} and ResNet20~\cite{he2016deep}. For ImageNet, we train VGG19 and MobileNetV1~\cite{howard2017mobilenets}. Model configurations and classification accuracy of validation set are shown in Table \ref{model_acc}.

\begin{table}[htbp]
\centering
\caption{Model configurations and accuracy.}
\resizebox{0.55\linewidth}{!}{
\begin{tabular}{ccrcrc}
\toprule 
\textbf{Datasets} & \textbf{Models} & \textbf{\#Parameters} & \textbf{\#Layers} & \textbf{\#Neurons} & \textbf{Accuracy} \\ \hline
\multirow{2}{*}{CIFAR-10} & VGG19        & 39,002,738  & 64  & 50,782 & 91.04\% \\
                          & AlexNet      & 87,650,186  & 12  & 18,282 & 91.24\% \\ \hline                     
\multirow{2}{*}{GTSRB}    & LeNet-5      & 172,331     & 7   & 291    & 98.26\% \\
                          & ResNet20     & 276,587     & 70  & 2,603  & 98.83\% \\ \hline
\multirow{2}{*}{ImageNet} & VGG19        & 139,666,034 & 64  & 50,782 & 95.40\% \\
                          & MobileNetV1  & 3,238,986   & 81  & 33,802 & 98.16\% \\  \bottomrule 
\label{model_acc}                        
\end{tabular}}
\end{table}

\textbf{Attacks:} We use following attack algorithms to generate adversarial examples of different perturbation sizes and distributions: FGSM~\cite{GoodfellowSS14}, BIM~\cite{Kurakin2017Adversarial}, JSMA~\cite{papernot2016limitations}, PGD~\cite{Madry2018Towards} as white-box attacks and AUNA~\cite{FoolboxAUNA}, PWA~\cite{SchottRBB19}, Boundary~\cite{Brendel2018Boundary} as black-box attacks. Besides, we also conduct AutoAttack (Auto)~\cite{croce2020reliable} for evaluation. The implementations of FGSM, BIM, JSMA, PGD, AUNA, PWA, Boundary are from the foolbox tool\footnote{The code of attacks can be downloaded at: https://
foolbox.readthedocs.io/en/v2.3.0/modules/attacks.html}. AutoAttack is conducted on adversarial robustness toolbox\footnote{https://github.com/Trusted-AI/adversarial-robustness-toolbox/tree/main/art}. Parameter settings of adversarial examples and attack success rate are shown in Table \ref{tab:Parameter_Setting_attack} and Table \ref{tab:ASR}, respectively. ``epsilon'' is the maximum perturbation size of the given example and ``stepsize'' is the perturbation step size of one iteration.

\begin{table}[htbp]
    \centering
    \caption{The attack parameter settings of adversarial attacks.
    }
    \label{tab:Parameter_Setting_attack}
    \resizebox{0.85\linewidth}{!}{
        \begin{tabular}{ll}
        \toprule 
        \multicolumn{1}{l}{\textbf{Attacks}} & \multicolumn{1}{l}{\textbf{Parameter Setting}}                                     \\ \hline
        FGSM                                 & $l_\infty$ epsilon=0.2, max\_iteration=1000.                                                      \\
        BIM                         & $l_\infty$ epsilon=0.3, stepsize=0.05, iterations=10. \\
        JSMA                               & $l_0$ epsilon=0.1, max\_iter=2000,max\_perturbations\_per\_pixel=7.                                                     \\
        PGD                                  & $l_\infty$ epsilon=0.3, stepsize=0.01,  iterations=40.     \\
        
        AUNA                           & $l_2$ epsilon=0.2, max\_iter=1000.                                                     \\
        PWA                                 & $l_2$ epsilon=0.3, max\_iter=1000.                                         \\
        Boundary                             & $l_2$ epsilon=0.2, iterations=5,000, max\_directions=25, initial\_stepsize=1e-2.\\
        Auto                           &  FGSM($l_\infty$ epsilon=0.2), PGD($l_\infty$ epsilon=0.2, stepsize=0.1), UAP($l_\infty$ epsilon=0.1, max\_iter=3). \\
         \bottomrule
        \end{tabular}
    } 
\end{table}

\begin{table}[htbp]
    \centering
    \caption{ASR of adversarial examples.
    }
    \label{tab:ASR}
    \resizebox{0.85\linewidth}{!}{
        \begin{tabular}{cccccccccc}
\toprule
\multirow{2}{*}{\textbf{Datasets}} & \multirow{2}{*}{\textbf{Models}} & \multicolumn{8}{c}{\textbf{Attacks}}                                                      \\ \cline{3-10} 
                          &                         & FGSM     & BIM     & JSMA    & PGD     & AUNA     & PWA     & Boundary & Auto    \\ \hline
\multirow{2}{*}{CIFAR-10} & VGG19   & 100.00\% & 99.85\% & 100.00\% & 99.90\% & 100.00\% & 97.60\% & 93.00\% & 84.50\% \\
                          & AlexNet                 & 99.55\%  & 96.20\% & 99.85\% & 95.90\% & 99.90\%  & 87.30\% & 95.60\%  & 70.80\% \\ \hline
\multirow{2}{*}{GTSRB}    & LeNet-5 & 99.60\%  & 97.55\% & 99.65\%  & 99.95\% & 99.95\%  & 99.85\% & 85.80\% & 75.95\% \\
                          & ResNet20                & 100.00\% & 99.95\% & 99.65\% & 99.90\% & 99.95\%  & 99.65\% & 83.40\%  & 82.75\% \\ \hline
\multirow{2}{*}{ImageNet} & VGG19   & 99.95\%  & 99.95\% & 99.85\%  & 99.90\% & 99.95\%  & 87.75\% & 87.50\% & 69.25\% \\
                          & MobileNetV1             & 100.00\% & 99.95\% & 99.95\% & 99.95\% & 100.00\% & 90.00\% & 89.25\%  & 78.75\% \\ \bottomrule
\end{tabular}
    } 
\end{table}

\textbf{Detection Baselines:} The following detection methods are adopted for comparison, including ANR~\cite{LiangLSLSW21}, LID~\cite{Ma0WEWSSHB18}, NNIF~\cite{CohenSG20}, NIC~\cite{ma2019nic} and ML-LOO~\cite{yang2020ml}. They are all recently-proposed detection algorithms, representing the SOTA results. We obtained the implementation of ANR, LID, NNIF and ML-LOO from GitHub\footnote{https://github.com/OwenSec/DeepDetector} \footnote{https://github.com/xingjunm/lid\_adversarial\_subspace\_detection} \footnote{https://github.com/giladcohen/NNIF\_adv\_defense}\footnote{https://github.com/Jianbo-Lab/ML-LOO}. We implement our copy of NIC following the introduced technical approach. As for parameters, we set $k$=10 and batch\_size=200 for calculating LID. For NNIF, $M$=10 and $C$ of logistic regression classifier is $1e5$. For ML-LOO, we choose last three fully-connected layers for feature extraction. All baselines are configured according to the best performance setting reported in the respective papers. 

\textbf{Evaluation Metrics:} The metrics used in the experiments are detailed as follows:

\textcircled{1} Classification accuracy: $acc=\frac{N_{true}}{N_{benign}}$, where $N_{true}$ is the number of benign examples correctly classified by the targeted model and $N_{benign}$ denotes the total number of benign examples.

\textcircled{2} Perturbation $l_2$-norm: $\rho=||x_{adv}-x||_2$, where $x$ and $x_{adv}$ are benign and its corresponding adversarial example respectively, and $||\cdot||_{2}$ represents $l_2$-norm. 

\textcircled{3} Attack success rate: $ASR = \frac{N_{adv}}{N_{benign}}$, where $N_{adv}$ denotes the number of adversarial examples.

\textcircled{4} Detection rate: $DR = \frac{N_{det}}{N_{mis}}$, where $N_{det}$ denotes the number of misclassified examples detected by detection methods.

\textcircled{5} Area under Curve (AUC) score~\cite{fawcett2006introduction}: The area under Receiver Operating Characteristic Curve. The higher, the better.

\textbf{Implementation Details:} (1) In our experiment, specific settings about training AdvD for each model is presented in Table~\ref{tab:advd}, unless otherwise specified. Misclassified perturbed examples are added with random noise to calculate local gradient. Structures of AdvD are all based on fully connected layers. (2) In the default setting, global average pooling is chosen for MobileNetV1, while flatten layer is chosen for other models. (3) We normalize the range of each pixel in the input images to [0, 1]. (4) All experiments are conducted on a server under Ubuntu 20.04 operating system with Intel Xeon Gold 5218R CPU running at 2.10GHz, 64 GB DDR4 memory, 4TB HDD and one GeForce RTX 3090 GPU card.

\begin{table}[htbp]
    \centering
    \caption{Details about training AdvD.
    }
    \label{tab:advd}
    \resizebox{0.95\linewidth}{!}{
        \begin{tabular}{ccccccccc}
\toprule
\multirow{2}{*}{\textbf{Dataset}} &
  \multirow{2}{*}{\textbf{Model}} &
  \multirow{2}{*}{\textbf{Chosen Layer}} &
  \multirow{2}{*}{\textbf{Input Shape}} &
  \multirow{2}{*}{\textbf{AdvD Structure}} &
  \multicolumn{2}{c}{\textbf{\#Training Examples}} &
  \multirow{2}{*}{\textbf{Batch Size}} &
  \multirow{2}{*}{\textbf{Epoch}} \\ \cline{6-7}
                          &             &                        &       &                      & Benign & Misclassified &    &    \\ \hline
\multirow{2}{*}{CIFAR-10} & VGG19       & flatten                & 512   & {[}512, 2{]}          & 10     & 200           & 32 & 7  \\
                          & AlexNet     & flatten                & 16384 & {[}512, 2{]}          & 10     & 200           & 32 & 7  \\ \hline
\multirow{2}{*}{GTSRB}    & LeNet-5     & flatten                & 1296  & {[}512, 512, 2{]}      & 15     & 200           & 32 & 15 \\
                          & ResNet20    & flatten                & 64    & {[}512, 512, 2{]}      & 15     & 200           & 32 & 15 \\ \hline
\multirow{2}{*}{ImageNet} & VGG19       & flatten                & 25088 & {[}1024, 512, 512, 2{]} & 10     & 300           & 64 & 7  \\
                          & MobileNetV1 & global average pooling & 1024  & {[}1024, 512, 512, 2{]} & 10     & 300           & 64 & 7  \\ \bottomrule
\end{tabular}
    } 
\end{table}

\subsection{Detection Against Adversarial Attacks}
We focus on the detection results of AdvCheck to measure the effectiveness against various adversarial examples. 

\begin{figure}[htbp]
\centering
    \subfigure[CIFAR-10, VGG19]{
        \includegraphics[width=0.32\linewidth]{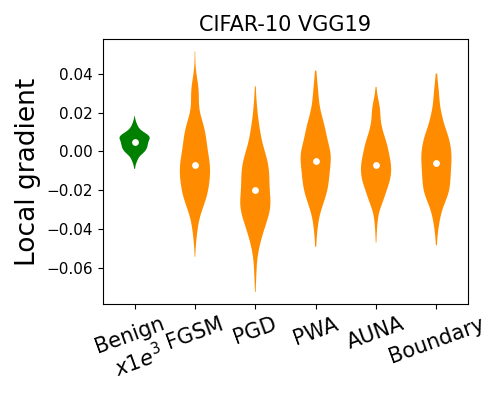}}
    \subfigure[CIFAR-10, AlexNet]{
        \includegraphics[width=0.32\linewidth]{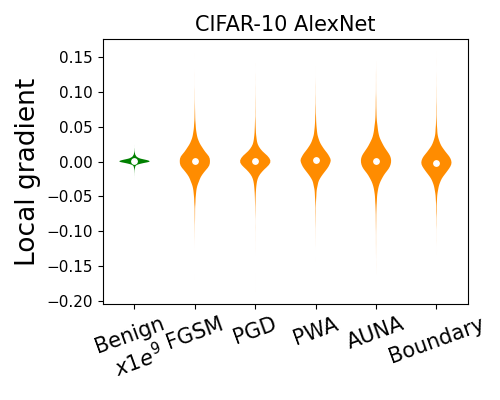}}
    \subfigure[GTSRB, LeNet-5]{
        \includegraphics[width=0.32\linewidth]{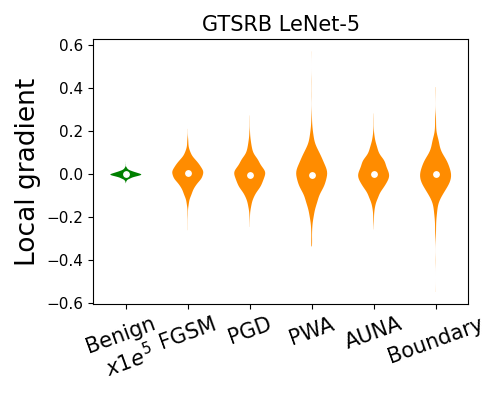}} \\
     \subfigure[GTSRB, ResNet20]{
        \includegraphics[width=0.32\linewidth]{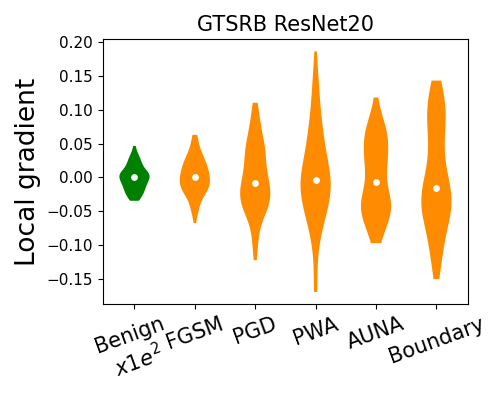}}
    \subfigure[ImageNet, VGG19]{
        \includegraphics[width=0.32\linewidth]{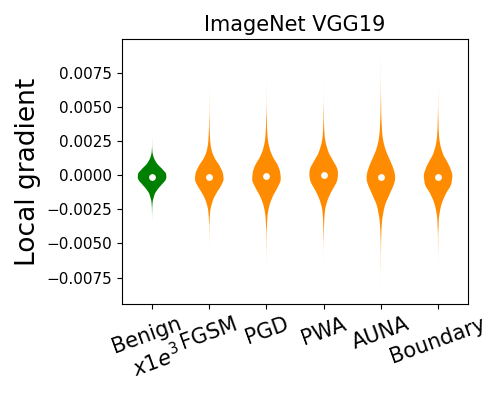}}
    \subfigure[ImageNet, MobileNetV1]{
        \includegraphics[width=0.32\linewidth]{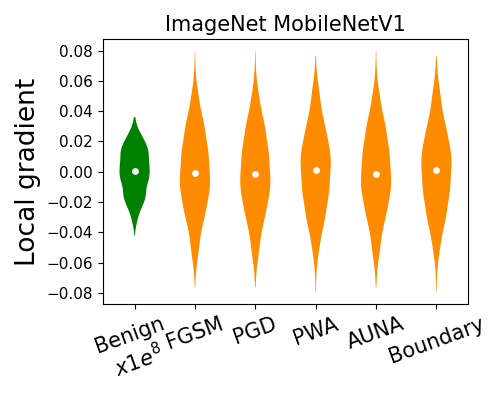}}   
\caption{Visualizations of local gradient of each model. Benign examples are colored in green while adversarial examples are painted in orange. }
\label{local_visual_all}
\end{figure}

\textbf{Implementation Details.} (1) 2,000 adversarial examples per attack are generated. Note that these adversarial examples are all generated from the correctly classified images from the training set. (2) We calculate DR against various adversarial attacks for measurement and compare AdvCheck with five recently-proposed baselines on six models of different datasets. Results are shown in Table \ref{detection}, where ``Benign" means detection accuracy on benign examples. Besides, the p-value of t-test between AdvCheck and each baseline is also calculated, shown in Table \ref{pvalue}. Two-tail and paired-sample t-test is adopted. p-value larger than 0.05 is bold. (3) We mix 2,000 adversarial examples per attack and 2,000 benign examples, and then calculate AUC score for each detection. Under this setting, VGG19 of CIFAR-10, LeNet-5 of GTSRB and MobileNetV1 of ImageNet are adopted. Results are shown in Table \ref{auc}. (4) We visualize the distribution of local gradient of six models in Fig. \ref{local_visual_all}. 

\begin{table*}[htbp]
\centering
\caption{Comparison between AdvCheck and baselines on Detecting adversarial examples. }
\resizebox{1\linewidth}{!}{
\begin{tabular}{cccccccccccc}
\toprule
\multirow{2}{*}{\textbf{Dataset}} &
  \multirow{2}{*}{\textbf{Model}} &
  \multirow{2}{*}{\textbf{Method}} &
  \multirow{2}{*}{\textbf{Benign}} &
  \multicolumn{8}{c}{\textbf{Attack}} \\ \cline{5-12} 
 &
   &
   &
   &
  FGSM &
  BIM &
  JSMA &
  PGD &
  AUNA &
  PWA &
  Boundary &
  Auto \\ \hline
\multirow{12}{*}{CIFAR-10} &
  \multirow{6}{*}{VGG19} &
  ANR &
  92.18\% &
  71.50\% &
  88.15\% &
  82.70\% &
  88.15\% &
  52.20\% &
  74.60\% &
  84.75\% &
  51.50\% \\
 &
   &
  LID &
  72.65\% &
  95.75\% &
  98.95\% &
  98.35\% &
  98.15\% &
  91.40\% &
  98.45\% &
  98.80\% &
  79.20\% \\
 &
   &
  NIC &
  98.95\% &
  93.00\% &
  89.15\% &
  90.50\% &
  89.10\% &
  90.36\% &
  90.50\% &
  98.30\% &
  93.20\% \\
 &
   &
  NNIF &
  99.00\% &
  94.00\% &
  92.00\% &
  96.00\% &
  88.00\% &
  96.00\% &
  94.00\% &
  86.00\% &
  92.40\% \\
 &
   &
  ML-LOO &
  86.32\% &
  85.40\% &
  88.70\% &
  89.00\% &
  87.70\% &
  83.60\% &
  84.00\% &
  86.30\% &
  95.70\% \\
 &
   &
  AdvCheck &
  \textbf{100.00\%} &
  \textbf{98.15\%} &
  \textbf{99.95\%} &
  \textbf{100.00\%} &
  \textbf{99.85\%} &
  \textbf{98.50\%} &
  \textbf{100.00\%} &
  \textbf{100.00\%} &
  \textbf{100.00\%} \\ \cline{3-12} 
 &
  \multirow{6}{*}{AlexNet} &
  ANR &
  92.40\% &
  66.75\% &
  80.75\% &
  77.50\% &
  83.25\% &
  40.75\% &
  78.00\% &
  26.25\% &
  23.25\% \\
 &
   &
  LID &
  91.52\% &
  84.30\% &
  95.40\% &
  94.60\% &
  83.60\% &
  83.30\% &
  97.20\% &
  86.70\% &
  79.20\% \\
 &
   &
  NIC &
  97.65\% &
  93.60\% &
  94.60\% &
  88.40\% &
  89.20\% &
  92.60\% &
  90.20\% &
  92.60\% &
  94.40\% \\
 &
   &
  NNIF &
  94.80\% &
  82.60\% &
  90.00\% &
  86.40\% &
  88.00\% &
  84.20\% &
  89.40\% &
  93.80\% &
  90.20\% \\
 &
   &
  ML-LOO &
  93.24\% &
  75.40\% &
  93.80\% &
  93.00\% &
  92.60\% &
  91.80\% &
  98.00\% &
  94.20\% &
  91.80\% \\
 &
   &
  AdvCheck &
  \textbf{99.47\%} &
  \textbf{100.00\%} &
  \textbf{99.00\%} &
  \textbf{100.00\%} &
  \textbf{99.20\%} &
  \textbf{100.00\%} &
  \textbf{100.00\%} &
  \textbf{100.00\%} &
  \textbf{99.70\%} \\ \hline
\multirow{12}{*}{GTSRB} &
  \multirow{6}{*}{LeNet-5} &
  ANR &
  90.60\% &
  59.60\% &
  70.20\% &
  78.00\% &
  76.60\% &
  17.40\% &
  46.75\% &
  30.00\% &
  22.00\% \\
 &
   &
  LID &
  51.45\% &
  70.20\% &
  91.00\% &
  72.40\% &
  91.35\% &
  72.65\% &
  51.25\% &
  51.30\% &
  71.50\% \\
 &
   &
  NIC &
  93.80\% &
  91.15\% &
  92.55\% &
  93.11\% &
  92.05\% &
  92.70\% &
  96.00\% &
  95.90\% &
  87.40\% \\
 &
   &
  NNIF &
  82.15\% &
  95.20\% &
  95.35\% &
  89.50\% &
  90.70\% &
  86.25\% &
  89.55\% &
  88.35\% &
  80.20\% \\
 &
   &
  ML-LOO &
  82.14\% &
  88.60\% &
  89.40\% &
  87.70\% &
  87.20\% &
  78.80\% &
  87.70\% &
  91.20\% &
  82.60\% \\
 &
   &
  AdvCheck &
  \textbf{99.35\%} &
  \textbf{99.70\%} &
  \textbf{99.55\%} &
  \textbf{99.60\%} &
  \textbf{99.65\%} &
  \textbf{95.55\%} &
  \textbf{98.85\%} &
  \textbf{98.85\%} &
  \textbf{99.60\%} \\ \cline{3-12} 
 &
  \multirow{6}{*}{ResNet20} &
  ANR &
  89.20\% &
  65.25\% &
  84.50\% &
  78.00\% &
  84.25\% &
  21.75\% &
  43.75\% &
  22.50\% &
  39.00\% \\
 &
   &
  LID &
  74.80\% &
  82.00\% &
  88.20\% &
  81.20\% &
  76.80\% &
  76.60\% &
  49.20\% &
  49.50\% &
  78.40\% \\
 &
   &
  NIC &
  92.80\% &
  91.50\% &
  92.40\% &
  92.80\% &
  93.80\% &
  72.60\% &
  89.20\% &
  88.40\% &
  75.40\% \\
 &
   &
  NNIF &
  82.40\% &
  87.60\% &
  86.40\% &
  86.00\% &
  84.80\% &
  70.60\% &
  90.20\% &
  85.40\% &
  69.00\% \\
 &
   &
  ML-LOO &
  88.36\% &
  83.20\% &
  82.60\% &
  90.40\% &
  88.20\% &
  78.20\% &
  91.40\% &
  80.60\% &
  75.30\% \\
 &
   &
  AdvCheck &
  \textbf{99.40\%} &
  \textbf{96.20\%} &
  \textbf{100.00\%} &
  \textbf{100.00\%} &
  \textbf{100.00\%} &
  \textbf{95.80\%} &
  \textbf{100.00\%} &
  \textbf{100.00\%} &
  \textbf{98.50\%} \\ \hline
\multirow{12}{*}{ImageNet} &
  \multirow{6}{*}{VGG19} &
  ANR &
  96.00\% &
  84.25\% &
  93.75\% &
  89.25\% &
  93.75\% &
  33.50\% &
  41.00\% &
  75.50\% &
  58.00\% \\
 &
   &
  LID &
  51.45\% &
  82.50\% &
  92.80\% &
  71.60\% &
  93.00\% &
  71.50\% &
  49.20\% &
  49.50\% &
  69.50\% \\
 &
   &
  NIC &
  90.60\% &
  94.50\% &
  95.40\% &
  96.20\% &
  93.60\% &
  94.40\% &
  96.70\% &
  96.80\% &
  95.10\% \\
 &
   &
  NNIF &
  75.60\% &
  77.40\% &
  88.00\% &
  82.40\% &
  88.80\% &
  88.00\% &
  66.00\% &
  76.60\% &
  92.40\% \\
 &
   &
  ML-LOO &
  85.00\% &
  88.50\% &
  84.10\% &
  85.20\% &
  92.60\% &
  78.40\% &
  84.20\% &
  89.60\% &
  78.40\% \\
 &
   &
  AdvCheck &
  \textbf{99.60\%} &
  \textbf{98.10\%} &
  \textbf{98.10\%} &
  \textbf{97.40\%} &
  \textbf{98.42\%} &
  \textbf{98.20\%} &
  \textbf{98.00\%} &
  \textbf{98.00\%} &
  \textbf{99.20\%} \\ \cline{3-12} 
 &
  \multirow{6}{*}{MobileNetV1} &
  ANR &
  85.60\% &
  83.20\% &
  88.20\% &
  86.00\% &
  88.20\% &
  48.60\% &
  46.50\% &
  78.75\% &
  57.75\% \\
 &
   &
  LID &
  92.90\% &
  83.95\% &
  87.03\% &
  87.15\% &
  87.00\% &
  64.95\% &
  66.00\% &
  66.35\% &
  81.20\% \\
 &
   &
  NIC &
  90.65\% &
  92.55\% &
  91.05\% &
  91.05\% &
  95.35\% &
  91.05\% &
  89.90\% &
  90.20\% &
  90.40\% \\
 &
   &
  NNIF &
  85.00\% &
  92.00\% &
  94.00\% &
  87.75\% &
  94.45\% &
  85.55\% &
  86.65\% &
  82.25\% &
  66.00\% \\
 &
   &
  ML-LOO &
  84.16\% &
  82.60\% &
  81.80\% &
  81.20\% &
  90.20\% &
  78.40\% &
  84.20\% &
  89.60\% &
  79.40\% \\
 &
   &
  AdvCheck &
  \textbf{95.00\%} &
  \textbf{96.23\%} &
  \textbf{96.35\%} &
  \textbf{96.55\%} &
  \textbf{96.35\%} &
  \textbf{92.85\%} &
  \textbf{97.00\%} &
  \textbf{98.15\%} &
  \textbf{99.50\%} \\ \bottomrule
  \label{detection}
\end{tabular}

}
\end{table*}

\begin{table}[htbp]
\centering
\caption{The p-value of t-test on DR between AdvCheck and each baseline. Values larger than 0.05 are bold.}
\resizebox{0.55\linewidth}{!}{
\begin{tabular}{ccccccc}
\toprule
\multirow{2}{*}{\textbf{Dataset}} & \multirow{2}{*}{\textbf{Model}} & \multicolumn{5}{c}{\textbf{Detection}}              \\ \cline{3-7} 
                          &             & ANR    & LID    & NIC    & NNIF   & ML-LOO \\ \hline
\multirow{2}{*}{CIFAR-10}         & VGG19                           & 0.0015 & \textbf{0.0605} & 0.0004 & 0.0018 & 0.0000 \\
                          & AlexNet     & 0.0032 & 0.0008 & 0.0001 & 0.0001 & 0.0051 \\ \hline
\multirow{2}{*}{GTSRB}    & LeNet-5     & 0.0009 & 0.0004 & 0.0003 & 0.0003 & 0.0000 \\
                          & ResNet20    & 0.0019 & 0.0007 & 0.0015 & 0.0001 & 0.0001 \\ \hline
\multirow{2}{*}{ImageNet} & VGG19       & 0.0154 & 0.0013 & 0.0026 & 0.0003 & 0.0000 \\
                          & MobileNetV1 & 0.0045 & 0.0017 & 0.0005 & 0.0118 & 0.0000 \\ \bottomrule
\end{tabular}}
\label{pvalue}
\end{table}

\begin{table}[htbp]
\centering
\caption{AUC scores between AdvCheck and baselines on detecting mixture examples. }
\resizebox{0.75\linewidth}{!}{
\begin{tabular}{cccccccccc}
\toprule
\multirow{2}{*}{\textbf{Dataset}} & \multirow{2}{*}{\textbf{Model}} & \multirow{2}{*}{\textbf{Method}} & \multicolumn{7}{c}{\textbf{Attack}} \\ \cline{4-10} 
 &
   &
   &
  FGSM &
  BIM &
  JSMA &
  PGD &
  AUNA &
  PWA &
  Boundary \\ \hline
\multirow{6}{*}{CIFAR-10} &
  \multirow{6}{*}{VGG19} &
  ANR &
  0.817 &
  0.901 &
  0.871 &
  0.905 &
  0.723 &
  0.835 &
  0.883 \\
 &
   &
  LID &
  0.924 &
  0.924 &
  0.988 &
  0.928 &
  0.900 &
  0.975 &
  0.982 \\
 &
   &
  NIC &
  0.947 &
  0.942 &
  0.923 &
  0.911 &
  0.901 &
  0.912 &
  0.971 \\
 &
   &
  NNIF &
  0.965 &
  0.955 &
  0.965 &
  0.935 &
  0.968 &
  0.965 &
  0.925 \\
 &
   &
  ML-LOO &
  0.935 &
  0.905 &
  0.924 &
  0.936 &
  0.918 &
  0.926 &
  0.931 \\
 &
   &
  AdvCheck &
  \textbf{0.998} &
  \textbf{0.998} &
  \textbf{0.998} &
  \textbf{0.998} &
  \textbf{0.995} &
  \textbf{0.998} &
  \textbf{0.998} \\ \hline
\multirow{6}{*}{GTSRB} &
  \multirow{6}{*}{LeNet-5} &
  ANR &
  0.751 &
  0.804 &
  0.843 &
  0.836 &
  0.540 &
  0.686 &
  0.603 \\
 &
   &
  LID &
  0.680 &
  0.928 &
  0.716 &
  0.930 &
  0.715 &
  0.492 &
  0.495 \\
 &
   &
  NIC &
  0.913 &
  0.943 &
  0.985 &
  0.982 &
  0.924 &
  0.974 &
  0.972 \\
 &
   &
  NNIF &
  0.937 &
  0.959 &
  0.877 &
  0.872 &
  0.887 &
  0.877 &
  0.912 \\
 &
   &
  ML-LOO &
  0.923 &
  0.927 &
  0.910 &
  0.896 &
  0.914 &
  0.937 &
  0.905 \\
 &
   &
  AdvCheck &
  \textbf{0.998} &
  \textbf{0.998} &
  \textbf{0.998} &
  \textbf{0.998} &
  \textbf{0.993} &
  \textbf{0.991} &
  \textbf{0.990} \\ \hline
\multirow{6}{*}{ImageNet} &
  \multirow{6}{*}{MobileNetV1} &
  ANR &
  0.844 &
  0.869 &
  0.858 &
  0.869 &
  0.671 &
  0.658 &
  0.821 \\
 &
   &
  LID &
  0.828 &
  0.907 &
  0.910 &
  0.906 &
  0.744 &
  0.769 &
  0.763 \\
 &
   &
  NIC &
  0.926 &
  0.926 &
  0.927 &
  0.978 &
  0.927 &
  0.907 &
  0.918 \\
 &
   &
  NNIF &
  0.925 &
  0.895 &
  0.852 &
  0.926 &
  0.863 &
  0.868 &
  0.853 \\
 &
   &
  ML-LOO &
  0.914 &
  0.924 &
  0.924 &
  0.93 &
  0.893 &
  0.923 &
  0.914 \\
 &
   &
  AdvCheck &
  \textbf{0.974} &
  \textbf{0.974} &
  \textbf{0.975} &
  \textbf{0.975} &
  \textbf{0.969} &
  \textbf{0.992} &
  \textbf{0.992} \\ \bottomrule
\end{tabular}
\label{auc}
}
\end{table}

\textbf{Results and Analysis.} In Table \ref{detection}, in all cases, AdvCheck shows superior detection (almost up to 100\%) on various adversarial attacks, when compared with baselines. And almost all p-values of NIP in Table \ref{pvalue} are smaller than 0.05. Besides, it outperforms both feature-based and model-based detection baselines with a considerable margin, especially on multi-class dataset like GTSRB. Specifically, on LeNet-5 of GTSRB, the DR of AdvCheck is 98.82\% on average, which is 1.8 times and 1.2 times that of ANR and NNIF. We speculate the reason that AdvCheck well captures the large difference between benign and adversarial examples on local gradient, which is easy to distinguish. 

Besides, AdvCheck shows quite stable detection results on different attacks, regardless of white-box attacks or black-box ones. When dealing with larger perturbations such as AUNA, DR and AUC don't show large fluctuations. This indicates the generality of AdvCheck. By learning the general pattern of attacks in local gradient, various adversarial examples can be detected. On the contrary, as for feature-based ANR, DR decreases when perturbations are larger.

When detecting benign examples, the classification accuracy of AdvCheck is the highest. This indicates that NIP hardly sacrifices the benign accuracy while completing effective detection. This is because local gradient of benign examples are much smaller than that of adversarial examples, consistent as shown in Figure \ref{local_visual_all}. Such large differences, independent on perturbation size and model structures, are easy for AdvCheck to distinguish. This also contributes to high AUC in mixture scenario in Table \ref{auc}. With regard to baselines, ANR removes adversarial perturbations together with important pixels, which decreases the accuracy on benign examples.

\begin{framed}
\textbf{Answer to RQ1:} AdvCheck outperforms the SOTA baselines in two aspects: (1) DR against various attacks ($\sim \times 1.3$ on average); (2) highest DR on benign examples; (3) higher AUC score ($\sim \times 1.2$ on average) on the mixture of benign and adversarial examples.  
\end{framed}

\subsection{Detection Against Misclassified Natural Inputs}
We further evaluate AdvCheck on two misclassification tasks, including misclassified examples in the training dataset and those due to weather conditions.

\begin{figure}[htbp]

    \makeatletter\subfigbottomskip=-13pt
\renewcommand{\@thesubfigure}{\hskip\subfiglabelskip}
\makeatother
    \centering
    \subfigure[]{
        \includegraphics[width=0.15\linewidth]{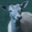}}
    \subfigure[]{
        \includegraphics[width=0.15\linewidth]{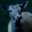}}
         \hspace{1.5pt} 
    \subfigure[]{
        \includegraphics[width=0.15\linewidth]{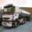}}
    \subfigure[]{
        \includegraphics[width=0.15\linewidth]{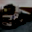}}
     \hspace{1.5pt} 
    \subfigure[]{
        \includegraphics[width=0.15\linewidth]{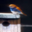}}
    \subfigure[]{
        \includegraphics[width=0.15\linewidth]{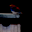}}
    \caption{Visualizations of weather examples.}
    \label{weather}
\end{figure}

\begin{figure}[htbp]
\centering
    \subfigure[Detection results]{
        \includegraphics[width=0.44\linewidth]{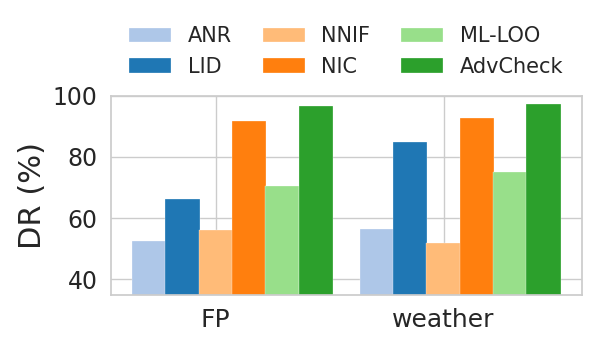}}
        \hspace{9pt}
    \subfigure[Visualizations of local gradient]{
        \includegraphics[width=0.42\linewidth]{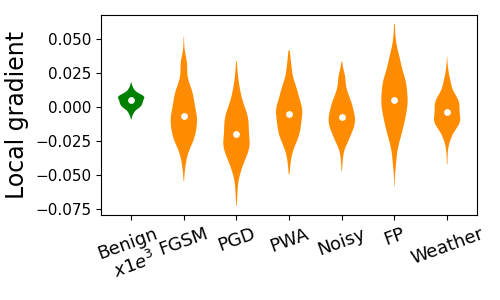}}
\caption{Detection performance and interpretable visualization.}
	\label{FP_weather}
\end{figure}

\textbf{Implementation Details.} (1) We conduct simple experiments on VGG19 of CIFAR-10. (2) 500 misclassified examples from training dataset (dubbed as ``FP'') are selected. For weather set-up, 500 misclassified examples are generated by DeepXplore \cite{pei2017deepxplore} to imitate natural weather conditions. Part of them are shown in Fig.\ref{weather}, where the left displays the original images while the right shows their corresponding weather examples. (3) The detection performance of AdvCheck, measured by DR, will be compared with baselines. Results are shown in Fig. \ref{FP_weather}(a). The distributions of local gradient of misclassified examples, including adversarial examples (FGSM, PGD, PWA) and misclassified natural inputs (Noisy, FP, Weather) are visualized in \ref{FP_weather}(b). The flatten layer in VGG19 of CIFAR-10 is used for visualization. Values of benign examples are magnified 1000 times.

\textbf{Results and Analysis.} As shown in Fig. \ref{FP_weather}(a), where deep green bars of AdvCheck is the highest among all detection methods. Specifically, DR of baselines is around 70\% on average against both misclassified inputs while DR of AdvCheck is up to 97\%. The outstanding performance of AdvCheck is mainly because the distribution learned by AdvCheck takes into account those misclassified examples, whose local gradient is similar to that activated by adversarial examples (consistent with results shown in Fig. \ref{FP_weather}(b)). Thus, AdvCheck's detection can be conducted more accurately on those natural examples. 

\begin{framed}
\textbf{Answer to RQ2:} Apart from detecting adversarial examples, AdvCheck can be applied to characterize misclassified natural inputs, e.g., FP and weather examples. Its performs better on DR ($\sim \times 1.4$ on average) than baselines. 
\end{framed}

\subsection{Potential Countermeasures}
If the detection is deterministic for the same adversarial example, the attacker can adaptively adopt specific methods against detection. We discuss potential adaptive attacks to fairly evaluate the effectiveness of AdvCheck. 

We have considered the adaptive attacks that fool the model while trying to minimize the difference of the local gradient with benign examples. Adversarial examples are calculated by:
\begin{equation}
    \mathop{\arg\min}\limits_{x_{adv}}~ MSE(\mu_i(x), \mu_i(x_{adv}))- \lambda\mathcal{J}(\theta,x_{adv},y)
\end{equation}
where $MSE(\cdot, \cdot)$ is the mean square error and $\lambda$ is the balancing parameter, which is set to 1 by default. $\mathcal{J}$ is the cross-entropy loss of the model with parameters $\theta$, adversarial example $x_{adv}$ and truth label $y$. We calculate local gradient in flatten layer or global average pooling for $\mu(x)$.

\begin{figure}[htbp]
\centering
        \subfigure[GTSRB, ResNet20]{
        \includegraphics[width=0.4\linewidth]{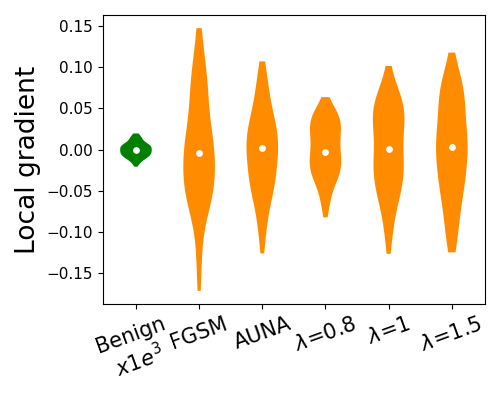}}
    \subfigure[ImageNet, MobileNetV1]{
        \includegraphics[width=0.4\linewidth]{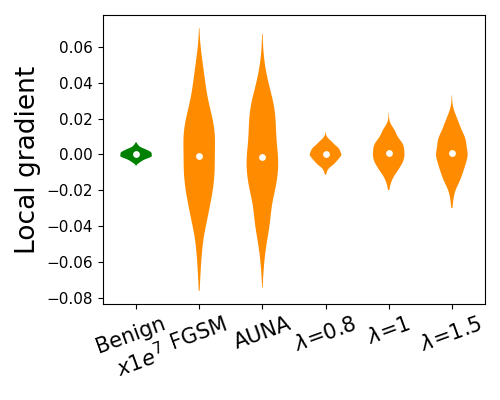}} 
\caption{Visualizations of local gradient for the adaptive attack. }
	\label{visual_adaptive}
\end{figure}

\textbf{Implementation Details.} (1) Experiments are conducted on ResNet20 of GTSRB and MobileNetV1 of ImageNet. (2) We iterated the calculation of each adversarial example for 10 rounds, because as iterations grow, ASR stays stable. (3) We set $\lambda$=0.8, 1, 1.5 and 2 to further evaluate the effectiveness of AdvCheck. (4) ASR of this adaptive attack is calculated on 1,000 adversarial examples. We count DR of AdvCheck for evaluation. Results are shown in Table \ref{tab:adaptive}. We also visualize the distribution of local gradient for the designed adaptive attack under different $\lambda$ in Fig. \ref{visual_adaptive}.

\begin{table}[htbp]
    \centering
    \caption{ASR and detection performance of AdvCheck on adaptive attacks.
    }
    \label{tab:adaptive}
    \resizebox{0.85\linewidth}{!}{
        \begin{tabular}{cccccc|cccc}
\toprule
\multirow{2}{*}{\textbf{Dataset}} & \multirow{2}{*}{\textbf{Model}} & \multicolumn{4}{c|}{\textbf{ASR}} & \multicolumn{4}{c}{\textbf{DR}} \\ \cline{3-10}  
         &             & $\lambda$=0.8 & $\lambda$=1 & $\lambda$=1.5 & $\lambda$=2 & $\lambda$=0.8 & $\lambda$=1 & $\lambda$=1.5 & $\lambda$=2 \\ \hline
GTSRB    & ResNet-20   & 84.20\%    & 84.50\%  & 88.00\%    & 92.10\%  & 90.90\%    & 96.30\%  & 95.80\%    & 96.40\%  \\
ImageNet & MobileNetV1 & 90.00\%    & 94.00\%  & 92.40\%    & 92.60\%  & 81.60\%    & 84.20\%  & 93.20\%    & 94.40\%  \\ \bottomrule
\end{tabular}
    } 
\end{table}

\textbf{Results and Analysis.} From Table \ref{tab:adaptive}, we can observe that most of these adaptive attacks are invalid against AdvCheck. This indicates that when faced with adaptive attacks, AdvCheck can still achieve high detection performance. Although local gradient of adversarial examples have similar distribution to those of benign examples, large difference in value bounds can still be observed from Fig. \ref{visual_adaptive}, which can be further used for AdvCheck to distinguish adversarial examples. 

Theoretically, larger $\lambda$ means smaller similarity in local gradient between benign and adversarial examples. So consistent with visualizations, local gradient of adversarial examples generated with larger $\lambda$ is more similar with that of general adversarial examples like FGSM. This makes it easier for AdvCheck to characterize adversarial properties. Thus, DR of AdvCheck increases.  

\begin{framed}
\textbf{Answer to RQ3:} AdvCheck shows reliability when facing potential countermeasures, i.e., it attains high defense efficacy (about 90\%).
\end{framed}

\subsection{Parameter Sensitivity Analysis}
According to the above discussion, the training examples and chosen layer will affect detection performance. Therefore, in this section, we analyze the effects on DR against various attacks.

\subsubsection{Impact of Training Examples}
In the default setting, AdvD is trained with benign examples and noise-added misclassified examples. We further investigate the effectiveness of AdvCheck, when it is trained with other perturbation-added examples. 

\begin{figure}[htbp]
\centering
        \subfigure[CIFAR-10, VGG19]{
        \includegraphics[width=0.45\linewidth]{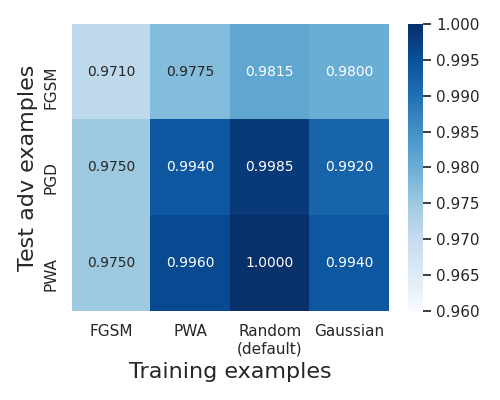}}
    \subfigure[GTSRB, ResNet20]{
        \includegraphics[width=0.45\linewidth]{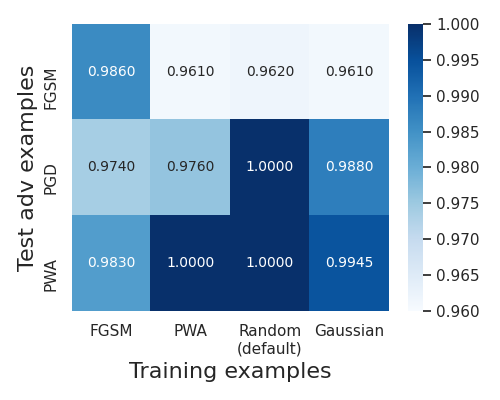}} 
\caption{The detection performance of AdvCheck when training with different examples. }
	\label{training}
\end{figure}

\textbf{Implementation Details.} (1) We study the impact of training examples on VGG19 of CIFAR-10 and ResNet20 of GTSRB. (2) For training AdvD, we use adversarial examples of FGSM and PWA, along with misclassified examples added by random noise (default setting) and Gaussian noise. Other training setting is same with that mentioned in Section \ref{setup}. (3) For measurement, 4,000 adversarial examples crafted by FGSM, PGD and PWA are used to calculate DR. The results are shown in Fig. \ref{training}. The horizontal line represents examples used for training, and the vertical line represents those for detection.

\textbf{Results and Analysis.} Little difference between DR among four training examples can be observed, with DR all over 97\%. This indicates that the detection performance of AdvCheck remains stable when various kinds of misclassified examples are used for training. The possible reason is that different perturbations that lead to wrong predictions have similar impact on local gradient, which is consistent with our assumption. AdvCheck performs a little better when trained with random noise than Gaussian noise. Generally speaking, as adopted in AdvCheck, generating misclassified examples from random noise is a more cost-saving way. Explicit prior knowledge of attacks is not required, making it more applicable to detect unknown attacks.

\subsubsection{Impact of Chosen Layer}
In this part, we change the layer of each targeted model for calculating local gradient to study the influence of the chosen layer.

\begin{figure}[htbp]
\centering
        \subfigure[CIFAR-10, VGG19]{
        \includegraphics[width=0.47\linewidth]{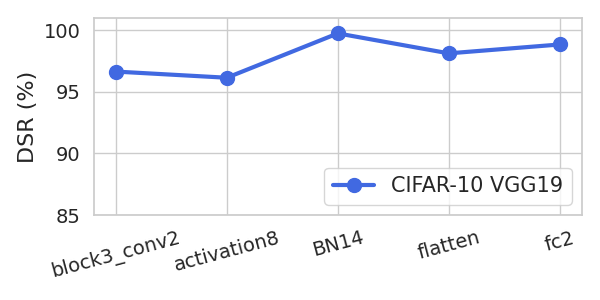}}
    \subfigure[GTSRB, ResNet20]{
        \includegraphics[width=0.47\linewidth]{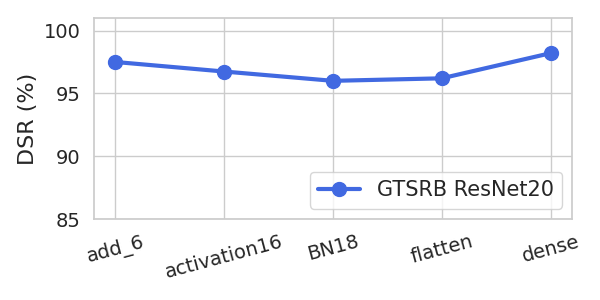}} 
\caption{The results of detection against FGSM under different chosen layer. ``BN'' denotes batch normalization.}
	\label{chosen_layer}
\end{figure}

\textbf{Implementation Details.} (1) VGG19 of CIFAR-10 and ResNet20 of GTSRB are adopted here. (2) We choose different layers from different types and depth (e.g., convolutional layer, batch normalization, fully-connected layer) to calculate local gradient from benign and misclassified examples. Then follow the default training parameters to train AdvD for detection. (3) We calculate DR on 4,000 adversarial examples crafted by FGSM for measurement. The results are shown in Fig. \ref{chosen_layer}, where the name of the chosen layer is shown in the x-axis.

\textbf{Results and Analysis.} It can be observed that two curves remain over 95\% as the chosen layer changes. Although DR of BN and fully-connected layer exceeds that of flatten, the detection result of AdvCheck is quite insensitive to different chosen layers, consistent with the empirical study shown in Section \ref{emp_study}. We speculate the reason that local gradients of benign and adversarial examples still show large difference, which is easy for detector to learn and characterize. This further demonstrates the effective design of AdvCheck.

\begin{framed}
\textbf{Answer to RQ4:} AdvCheck is quite insensitive when choosing different training examples and network layers, i.e., DR of it is still over 95\%. 
\end{framed}

\section{Conclusion and Discussion }
In this paper, we propose the concept of local gradient, and observe that benign and adversarial examples show quite large difference in the same layer in the DNN. Based on it, we design AdvCheck, a general framework for detecting adversarial examples and even misclassified natural inputs. It only relies on a few benign examples to achieve attack-agnostic detection. Extensive experiments have verified that compared with SOTA baselines. 

Although AdvCheck has demonstrated its effectiveness of defending various adversarial attacks, it assumes the access of the model and requires a few benign examples. There are situations where the layer output of the model and benign examples may not be available. It is unclear how our method can be extended to handle those cases. We will leave it to our future work. Besides, we will work towards expanding our work to other domains, such as object detection and natural language processing.


\section*{Acknowledgments}

This research was supported by the National Natural Science Foundation of China (No. 62072406), Zhejiang Provincial Natural Science Foundation (No. LDQ23F020001), Chinese National Key Laboratory of Science and Technology on Information System Security (No. 61421110502) and National Key R\&D Projects of China (No. 2018AAA0100801).

\vfill

\end{document}